\documentclass[10pt]{article}
\usepackage{amsmath}
\usepackage{amssymb}
\usepackage{graphicx}
\usepackage{cite}
\usepackage{color}
\usepackage[labelfont=bf,labelsep=period,justification=raggedright]{caption}

\makeatletter
\renewcommand{\@biblabel}[1]{\quad#1.}
\makeatother

\date{}

\begin{document}

\begin{flushleft}
{\Large
\textbf{Optimal distribution of incentives for public cooperation in heterogeneous interaction environments}
}\sffamily
\\[3mm]
\textbf{Xiaojie Chen$^{1,\ast}$, Matja{\v z} Perc$^{2,3,\dagger}$}
\\[2mm]
{\bf 1} School of Mathematical Sciences, University of Electronic Science and Technology of China,
Chengdu 611731, China, \,{\bf 2} Faculty of Natural Sciences and Mathematics, University of Maribor, Koro{\v s}ka cesta 160, SI-2000 Maribor, Slovenia, {\bf 3} CAMTP -- Center for Applied Mathematics and Theoretical Physics, University of Maribor, Krekova 2, SI-2000 Maribor, Slovenia\\
$^\ast$xiaojiechen@uestc.edu.cn, $^\dagger$matjaz.perc@uni-mb.si
\end{flushleft}
\sffamily

\section*{Abstract}
In the framework of evolutionary games with institutional reciprocity, limited incentives are at disposal for rewarding cooperators and punishing defectors. In the simplest case, it can be assumed that, depending on their strategies, all players receive equal incentives from the common pool. The question arises, however, what is the optimal distribution of institutional incentives? How should we best reward and punish individuals for cooperation to thrive? We study this problem for the public goods game on a scale-free network. We show that if the synergetic effects of group interactions are weak, the level of cooperation in the population can be maximized simply by adopting the simplest ``equal distribution'' scheme. If synergetic effects are strong, however, it is best to reward high-degree nodes more than low-degree nodes. These distribution schemes for institutional rewards are independent of payoff normalization. For institutional punishment, however, the same optimization problem is more complex, and its solution depends on whether absolute or degree-normalized payoffs are used. We find that degree-normalized payoffs require high-degree nodes be punished more lenient than low-degree nodes. Conversely, if absolute payoffs count, then high-degree nodes should be punished stronger than low-degree nodes.

\section*{Introduction}
In human societies, cooperation is essential for the maintenance of
public goods. However, the collapse of cooperation happens often in
many public goods dilemmas which we nowadays face, like protecting
the global climate or avoiding overfishing of our oceans
\cite{hardin_g_s68, o'neill_s02}. For avoiding the tragedy of the
commons \cite{hardin_g_s68}, we often rely on institutions to
enforce public cooperation and acceptable behavior. Although
institutionalized punishment appears to be more common than
institutionalized reward, both concepts are in use throughout the
world. Recently, ample research efforts have been devoted to the
study of the emergence of institutions and their effectiveness in
promoting prosocial behavior \cite{yamagishi_jpsp86, ostrom_90,
gurerk_s06, henrich_s06, cuesta_jtb08, sigmund_n10, sasaki_jtb11,
baldassarri_pnas11, szolnoki_pre11, szolnoki_pre11b, cressman_jtb12,
sasaki_pnas12, isakov_dga12, vasconcelos_ncc13, vukov_pcbi13,
bechtel_pnas13, cressman_bt13}. It has been shown, for example, that
institutional rewarding promotes the evolution of cooperation in the
liner public goods game \cite{cuesta_jtb08}, the nonlinear public
goods games \cite{chen_xj_jtb13}, and in structured populations in
general \cite{jimenez_jtb08, jimenez_r_jeic09, szolnoki_pre11,
szolnoki_pre11b}. However, institutional punishment is less costly
and thus more effective to warrant a given level of public
cooperation, especially if participation in the public goods game is
optional \cite{sasaki_pnas12, sasaki_dga13}.

Besides the obvious stick versus carrot dilemma \cite{hilbe_prsb10, szolnoki_njp12, szolnoki_prx13}, the question emerges how to best make use of the available resources, which inevitably are finite \cite{perc_srep12, chen_xj_sr14}. In particular, we wish to make optimal use of different forms of reciprocity to promote human cooperation \cite{poteete_10, gracia-lazaro_srep12, gracia-lazaro_pnas12, exadaktylos_srep13, rand_tcs13}. One plausible approach appears to be allocating the resources depending on the properties of the interaction network that describes the connections among us. Surprisingly, few studies have thus far considered the problem of the optimal allocation of incentives for maximizing public cooperation. Traditionally, all groups and all individuals are considered equal, and depending on their strategies thus deserved of the same reward or punishment \cite{jimenez_jtb08, sasaki_pnas12, chen_xj_jtb13}. This simple assumption, however, does not agree with the fact that in social networks individuals have different roles, which depend significantly on the degree of the node that they occupy. Indeed, the prominent role of heterogeneous interaction networks for the successful evolution of cooperation is firmly established and well known \cite{santos_prl05, santos_pnas06, santos_ploscb06, gomez-gardenes_prl07, masuda_prsb07, fu_pa07, tanimoto_pre07, tomassini_ijmpc07, fu_pre08b, assenza_pre08, santos_n08, floria_pre09, tanimoto_pa09, pena_pre09, fu_pre09, poncela_epl09, pacheco_ploscb09, brede_epl11, gomez-gardenes_epl11, tanimoto_pre12, pinheiro_pone12, kovarik_pa12, simko_pone13, tanimoto_pre13}, and the reasonable assumption is that the incentives should likely also be distributed accordingly for optimal evolutionary outcomes. In the framework of evolutionary graph theory the interaction groups are diverse, and naturally thus the provided incentives within each group should also be different. The number of links an individual player has is traditionally assumed to be a good proxy for that player's influence and importance. In this sense, it is interesting and highly relevant to determine how to distribute the incentives in the light of this heterogeneity.

Here we consider the spatial public goods game \cite{szolnoki_pre09c} with institutional reciprocity on a scale-free network \cite{barabasi_s99, albert_rmp02}, where the assumption is that the incentives at disposal for rewarding cooperators and punishing defectors are limited. We assume that the budget available to each group is proportional to its size, and that the distribution of the incentives depends on the number of individual links within the group. Our aim is to arrive at a thorough understanding of how the incentives should be best distributed to maximize public cooperation. In what follows, we present the results obtained with the model described in the Methods section, to where we refer for details. As we will show, if the enhancement factor $r$ is small, the level of cooperation can be maximized simply by adopting the simplest ``equal distribution'' scheme. If the value of $r$ is large, however, it is best to reward high-degree nodes more than low-degree nodes. Unlike for institutionalized rewards, the optimal distribution of resources within the framework of institutional punishment depends on whether absolute or degree-normalized payoffs are used. High-degree nodes should be punished more lenient than low-degree nodes if degree-normalized payoffs apply, while high-degree nodes should be punished stronger than low-degree nodes if absolute payoffs count.

\section*{Results}
We perform Monte Carlo simulations of the public goods game described in the Methods section, whereby we consider separately institutional rewarding with absolute payoffs and institutional punishment with absolute payoffs, as well as institutional rewarding with degree-normalized payoffs and institutional punishment with degree-normalized payoffs. As the key parameters we consider the enhancement factor $r$, the average amount of available incentives $\delta$, and the distribution strength of incentives $\alpha$ (see Methods for details). We determine the stationary fractions of cooperators in the stationary state on networks comprising $N=1000$ to $10000$ nodes. The final results are averaged over $100$ independent initial conditions to further enhance accuracy.

\subsection*{Institutional rewarding with absolute payoffs}
\begin{figure}
\begin{center}
\includegraphics[width=10cm]{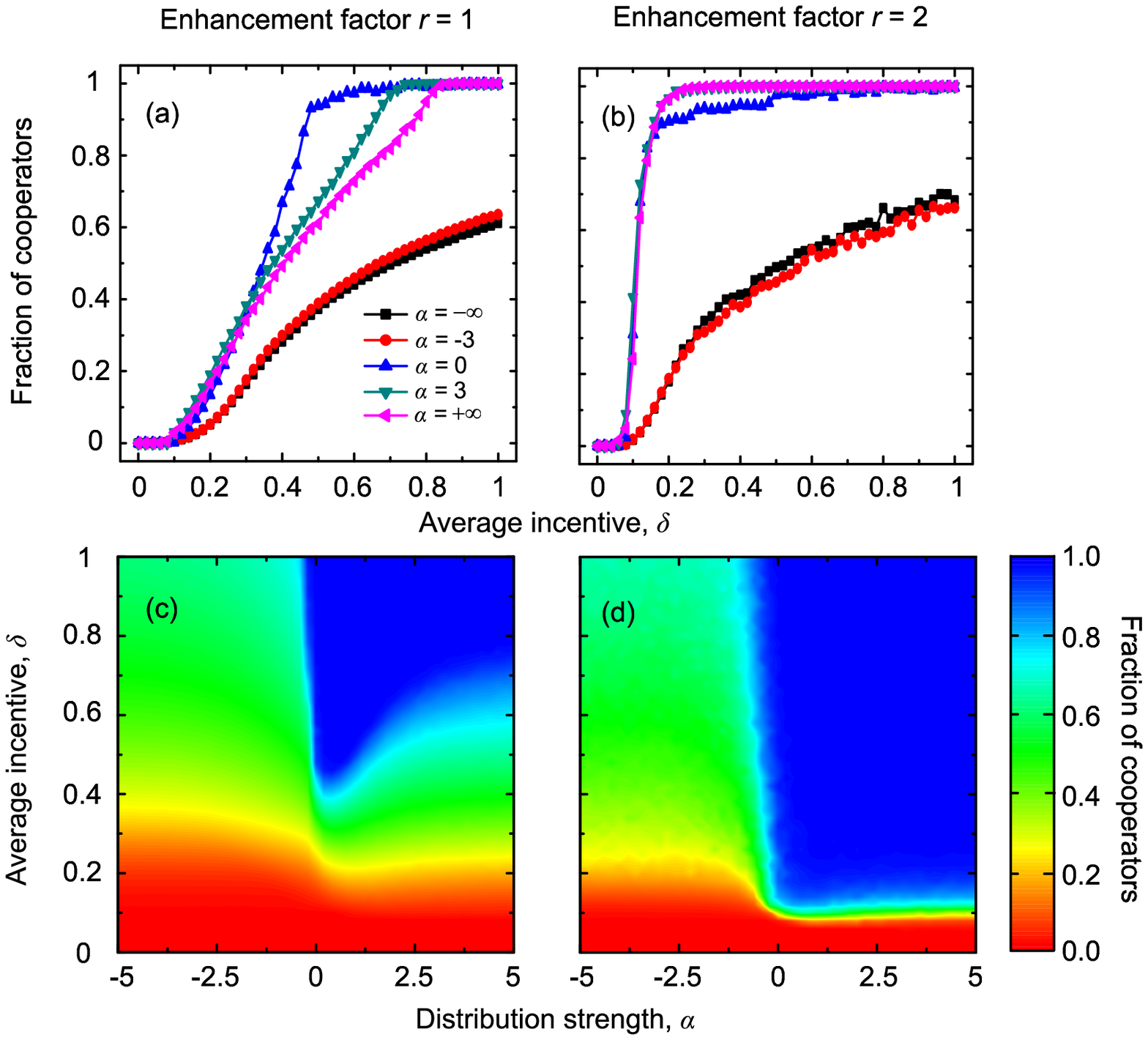}
\end{center}
\caption{Evolution of cooperation with institutional rewarding and absolute payoffs. Top row depicts the stationary fraction of cooperators as a function of the average amount of available incentives $\delta$ for different values of the distribution strength $\alpha$. The enhancement factor is $r=1$ in (a) and $r=2$ in (b). Bottom row depicts the contour plot of the fraction of cooperators as a function of $\alpha$ and $\delta$, as obtained for the enhancement factor $r=1$ in (c) and $r=2$ in (d).}
\label{fig1}
\end{figure}

Figure~\ref{fig1} shows the stationary fraction of cooperators for two different values of the enhancement factor $r$. When the enhancement factor is small [Fig.~\ref{fig1}(a) and (c)], defectors always dominate if $\delta <0.2$, and this regardless of the value of $\alpha$. For intermediate values of $\delta$, the cooperation level can be maximized at an intermediate value of the distribution strength $\alpha$, which ought to be close to zero. This indicates that an equal distribution of positive incentives, regardless of the degree of players within the group, is the optimal distribution scheme for public cooperation. For high values of $\delta$, the cooperation level increases with increasing the value of $\alpha$. If the enhancement factor $r$ increases
[Fig.~\ref{fig1}(b) and (d)], defectors still dominate for small values of $\delta$ and regardless of the value of $\alpha$. However, the nonmonotonous dependence of the fraction of cooperators on the distribution strength $\alpha$ disappears for intermediate values of $\delta$. Instead, the highest cooperation level is attainable for large values of $\alpha$.

Intuitively, it is possible to understand that when the enhancement factor is small, a modest positive incentive is not enough to reverse the doom of cooperators, no matter which distribution scheme is used \cite{sasaki_pnas12}. Conversely, if the incentives are large and targeted preferentially towards influential players, they can have a high payoff even if the part stemming from the public goods game is small. In agreement with the traditional argument of network reciprocity \cite{nowak_n92b, santos_prl05, wang_z_srep13}, only cooperators are able to forge a long-term advantage out of this favorable situation and build sizable cooperative groups. Thus, for high-enough values of $\alpha$, which favor the distribution of rewards towards high-degree nodes, the evolution of public cooperation is successful.

\begin{figure}
\begin{center}
\includegraphics[width=16cm]{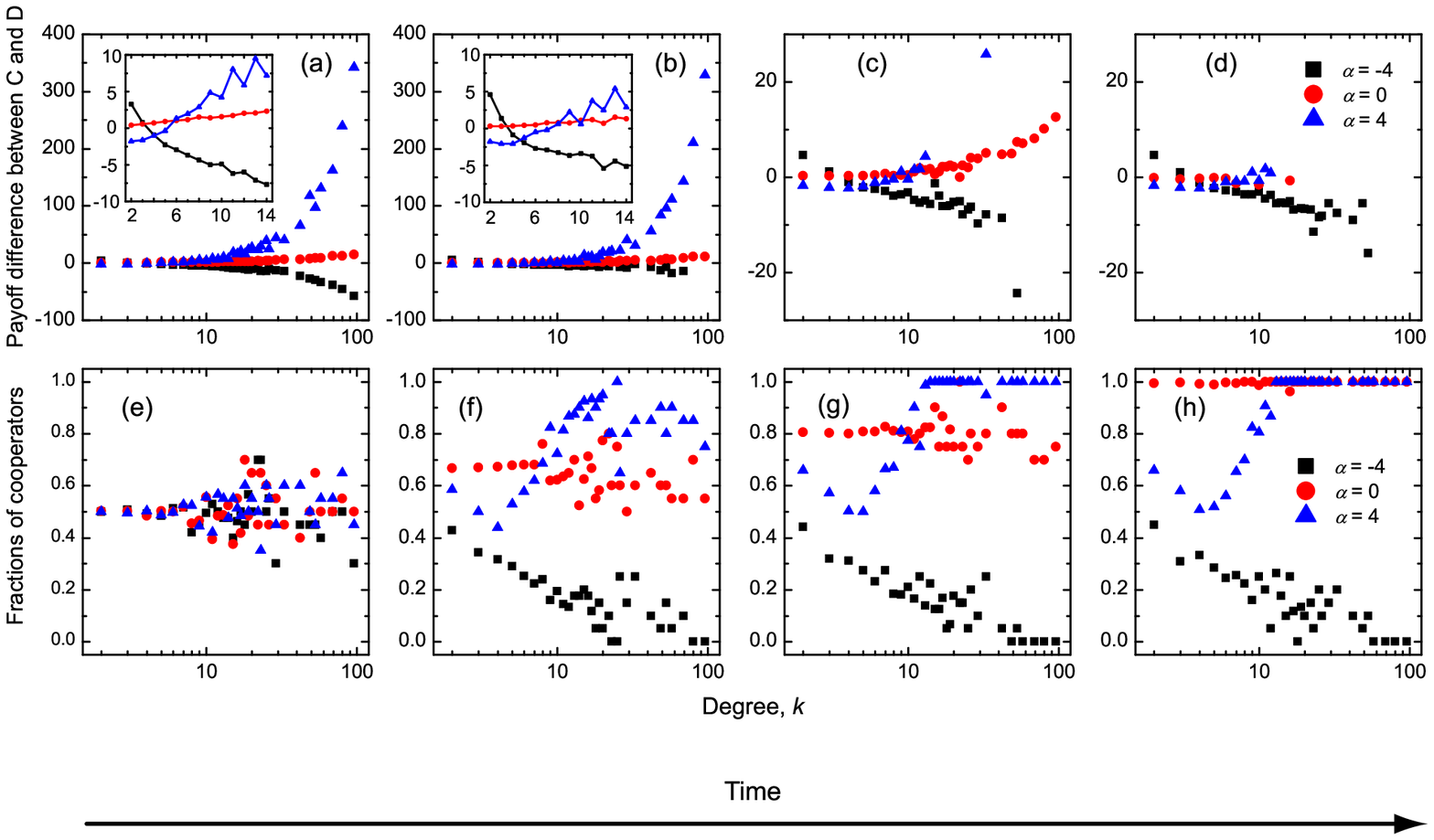}
\end{center}
\caption{Time evolution of the mean payoff difference between cooperators and defectors (top row) and the fraction of cooperators (bottom row) as a function of degree $k$ for three typical
values of $\alpha$. Institutional rewarding and absolute payoffs apply. The insets of (a) and (b) show the mean payoff difference between cooperators and defectors for low-degree and middle-degree nodes during the early stages of evolution. During the evolutionary process, if the enhancement factor is small, cooperators always have a higher mean payoff than defectors at an intermediate value of $\alpha$. Parameter values are $r=1$ and $\delta=0.5$.}
\label{fig2}
\end{figure}

To gain an understanding of the optimal intermediate value of the average amount of available incentives $\delta$ requires more effort. First, we show in Fig.~\ref{fig2} the payoff differences between cooperators and defectors as well as the fraction of cooperators as a function of degree $k$ during different typical evolutionary stages of the game. We observe that for $\alpha=0$, which implies an ``equal distribution'' scheme irrespective of the degree of players, the payoff of cooperators is higher than that of the corresponding defectors, and this regardless of $k$. Thus, cooperators can successfully occupy all nodes of the networks. In contrast, for negative values of $\alpha$, the payoff of cooperators with high or middle degree is less than that of the corresponding defectors, while cooperators with low degree have a higher mean payoff than defectors with small degree. Because there are interconnections among different types of nodes, and because the Fermi strategy updating rule is adopted, cooperators can coexist with defectors at equilibrium. But defectors can occupy most of the nodes in the network, since low-degree cooperators do not obtain a high enough payoff to spread their strategy across the network. For positive values of $\alpha$, the payoff of cooperators with high and middle degree is larger than that of the corresponding defectors, while cooperators with low degree have a lower mean payoff than defectors with low degree. In addition, high-degree cooperators obtain a sufficiently high payoff through institutional rewarding that enables them to spread the cooperative strategy also towards some of the auxiliary low-degree nodes. Accordingly, cooperative behavior prevails over defection, but the stationary state is still a mixed $C+D$ phase --- cooperators are unable to dominate completely.

\begin{figure}
\begin{center}
\includegraphics[width=16cm]{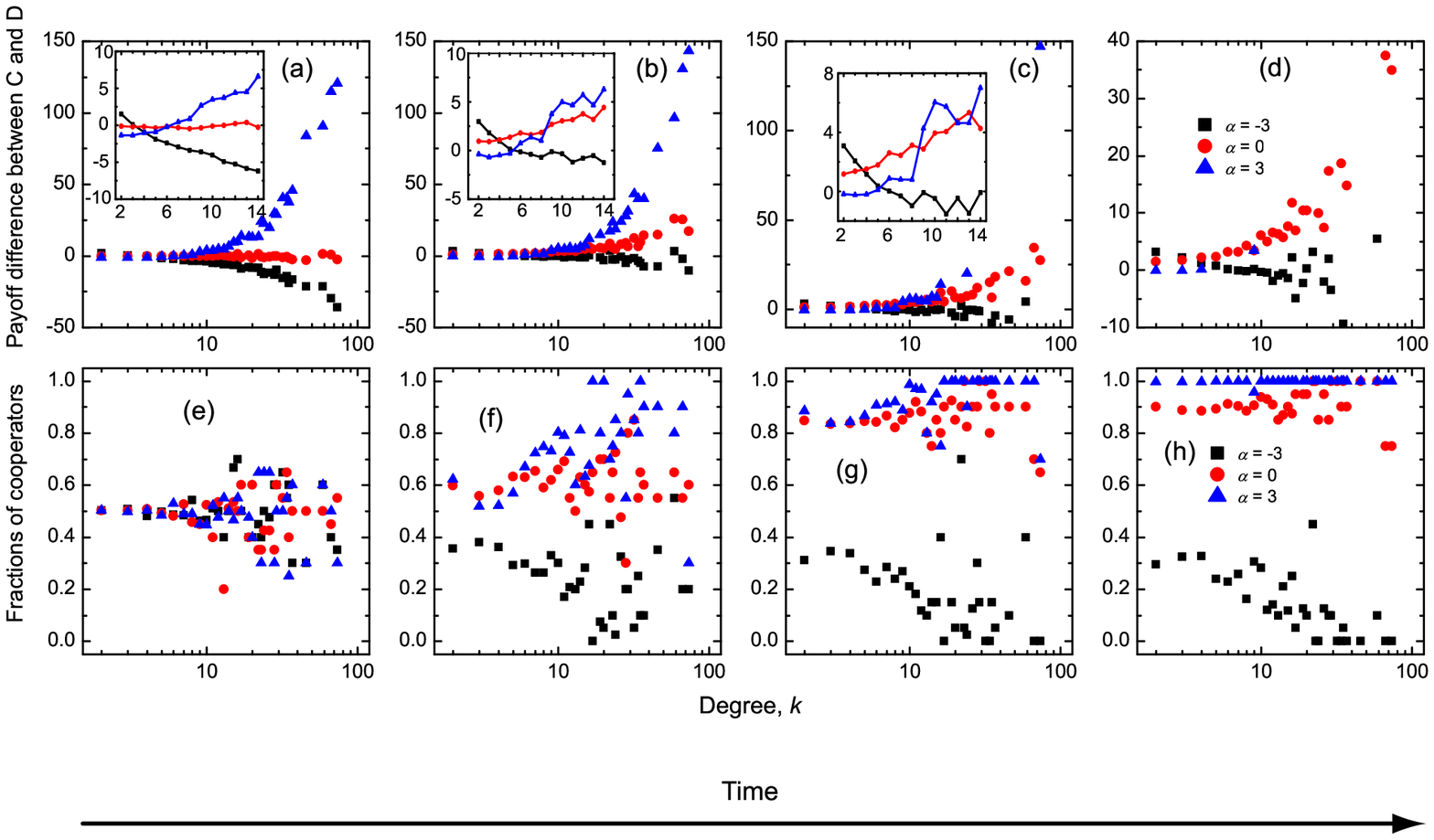}
\end{center}
\caption{The same results as presented in Fig.~\ref{fig2}, as obtained for a larger enhancement factor but a smaller average amount of available incentives. Parameter values are $r=2$ and $\delta=0.3$.}
\label{fig3}
\end{figure}

To further corroborate our arguments, we show in Fig.~\ref{fig3} the payoff difference between cooperators and defectors as well as the fraction of cooperators in dependence on
degree $k$, as obtained for a two times larger value of $r$ than used in Fig.~\ref{fig2}. In comparison, it can be observed that for $\alpha \leq 0$ the results remain unchanged. For $\alpha>0$, on the other hand, the process of evolution is different from what we have presented in Fig.~\ref{fig2}. During the early stages of evolution [Fig.~\ref{fig3} (a) and (b)], cooperators with low degree can have a lower mean payoff than low-degree defectors, while cooperators with middle and high degree can have a higher mean payoff than the corresponding defectors. Further in time, cooperators succeed in occupying all high-degree nodes [Fig.~\ref{fig3} (g)], and even low-degree cooperators have a payoff comparable to that of low-degree defectors [Fig.~\ref{fig3} (c)]. Cooperators can eventually invade the whole network [Fig.~\ref{fig3} (h)], thus giving rise to the absorbing $C$ phase at $r=2$, which emerges for sufficiently large values of $\alpha$ and intermediate values of $\delta$.

\subsection*{Institutional punishment with absolute payoffs}
\begin{figure}
\begin{center}
\includegraphics[width=10cm]{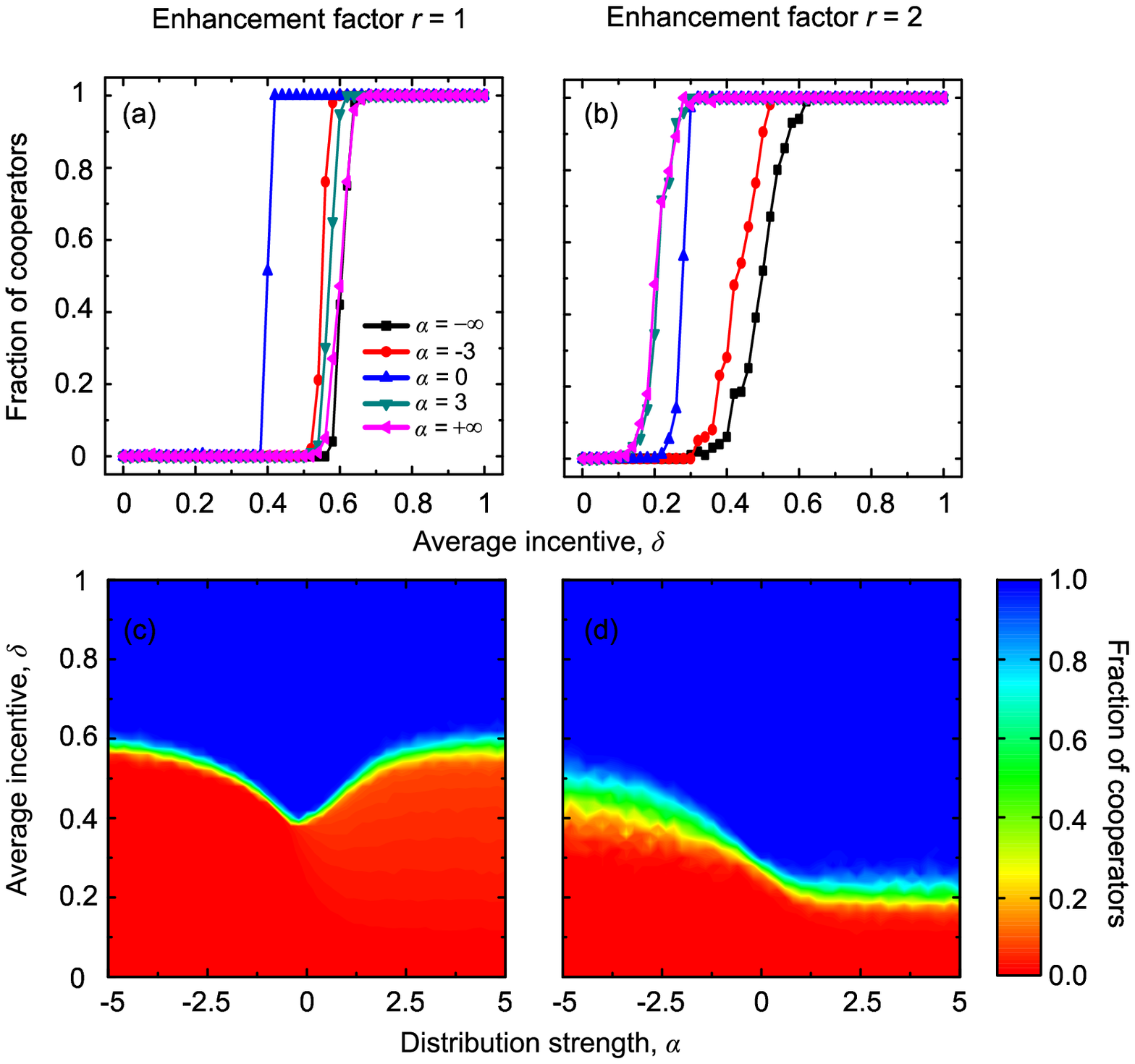}
\end{center}
\caption{Evolution of cooperation with institutional punishment and absolute payoffs. Top row depicts the stationary fraction of cooperators as a function of the average amount of available incentives $\delta$ for different values of the distribution strength $\alpha$. The enhancement factor is $r=1$ in (a) and $r=2$ in (b). Bottom row depicts the contour plot of the fraction of cooperators as a function of $\alpha$ and $\delta$, as obtained for the enhancement factor $r=1$ in (c) and $r=2$ in (d).}
\label{fig4}
\end{figure}

Figure~\ref{fig4} shows that when the enhancement factor is small, cooperators are unable to survive for small values of $\delta$, and this irrespective of the value of $\alpha$. For intermediate values of $\delta$, the highest cooperation level is attained at an intermediate value of the distribution strength $\alpha$, which is almost equal to zero, like by the consideration of institutional reward in the preceding subsection. This indicates that, for small enhancement factors, in case of institutional punishment too an ``equal distribution'' scheme works best for the evolution of public cooperation. If $\delta$ is large --- if resources for punishment abound --- cooperators can always dominate, regardless of the value of $\alpha$. For a two times larger value or $r$ cooperators are favored even more, so that the nonmonotonous dependence of the cooperation level on $\alpha$ at intermediate values of $\delta$ vanishes. Instead, the fraction of cooperators simply increases with increasing values of $\alpha$. Thus, the more the high-degree defectors are punished, the better for the evolution of public cooperation.

\begin{figure}
\begin{center}
\includegraphics[width=16cm]{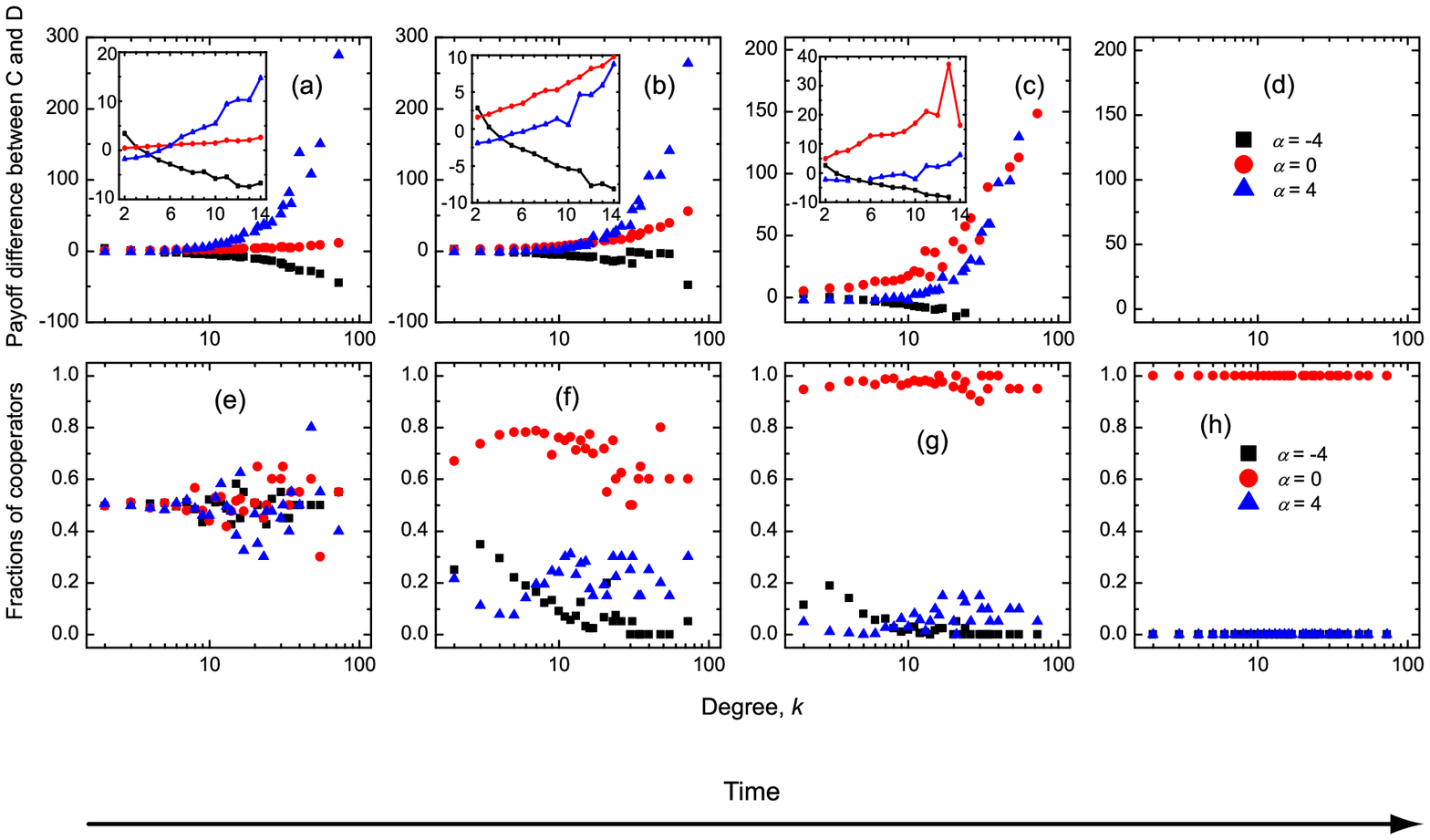}
\end{center}
\caption{Time evolution of the mean payoff difference between cooperators and defectors (top row) and the fraction of cooperators (bottom row) as a function of degree $k$ for three typical
values of $\alpha$. Institutional punishment and absolute payoffs apply. The insets of (a) and (b) show the mean payoff difference between cooperators and defectors for low-degree and middle-degree nodes during the early stages of evolution. During the evolutionary process, if the enhancement factor is small, cooperators always have a higher mean payoff than defectors at an intermediate value of $\delta$. Parameter values are $r=1$ and $\delta=0.5$.}
\label{fig5}
\end{figure}

It is understandable that low values of $r$, paired with modest resources for punishing defectors, lead to the dominance of defectors, regardless of the value of $\alpha$. Conversely, if the resources abound, defectors are punished severely and cooperators dominate. In this limit example the distribution of fines between low, middle and high degree nodes does not play an important role. If, however, the combination of values of $r$ and $\delta$ just barely, or not at all, support the survivability of cooperators, then the value of $\alpha$, and thus the particular distribution of incentives (in this case fines), plays a significant role. With the aim to explain this nontrivial dependence on $\alpha$, we show in Fig.~\ref{fig5} the payoff difference between cooperators and defectors as well as the fraction of cooperators as a function of degree $k$ during different stages of the evolutionary process. It can be observed that for $\alpha=0$ cooperators can always have a higher payoff than the defectors with the same corresponding degree [Fig.~\ref{fig5} (a)-(c)]. Cooperators can thus rise to complete dominance. While for $\alpha<0$, however, low-degree cooperators can have a higher mean payoff than low-degree defectors, while cooperators with middle or high degree can't match the corresponding defectors during the early stages of the evolution [Fig.~\ref{fig5} (a)]. Defectors can therefore, over time, occupy the high-degree and middle-degree nodes [Fig.~\ref{fig5} (f)]. This invasion can decrease the fraction of cooperators on low-degree nodes \cite{chen_xj_pla08}. Accordingly, the payoff difference between cooperators and defectors on these nodes continues to be negative, although low-degree
defectors receive the negative incentives. Ultimately cooperators therefore die out. For $\alpha>0$, on the other hand, defectors with higher degree are punished preferentially --- they receive a bigger share of fines from the available fond than low-degree defectors.
Due to the small enhancement factor and the institutional punishment, both high-degree cooperators and high-degree defectors have negative payoffs. In fact, low-degree players can have a higher payoff than high-degree players, despite of the fact that we use absolute payoffs in this particular case. Either way, defectors can easily invade low-degree nodes [Fig.~\ref{fig5}(f)],
and they can spread further towards middle and high degree nodes, although at the beginning of evolution cooperators have a higher payoff than defectors on these nodes. The ultimate consequence is that defectors dominate completely [Fig.~\ref{fig5} (h)].

\begin{figure}
\begin{center}
\includegraphics[width=16cm]{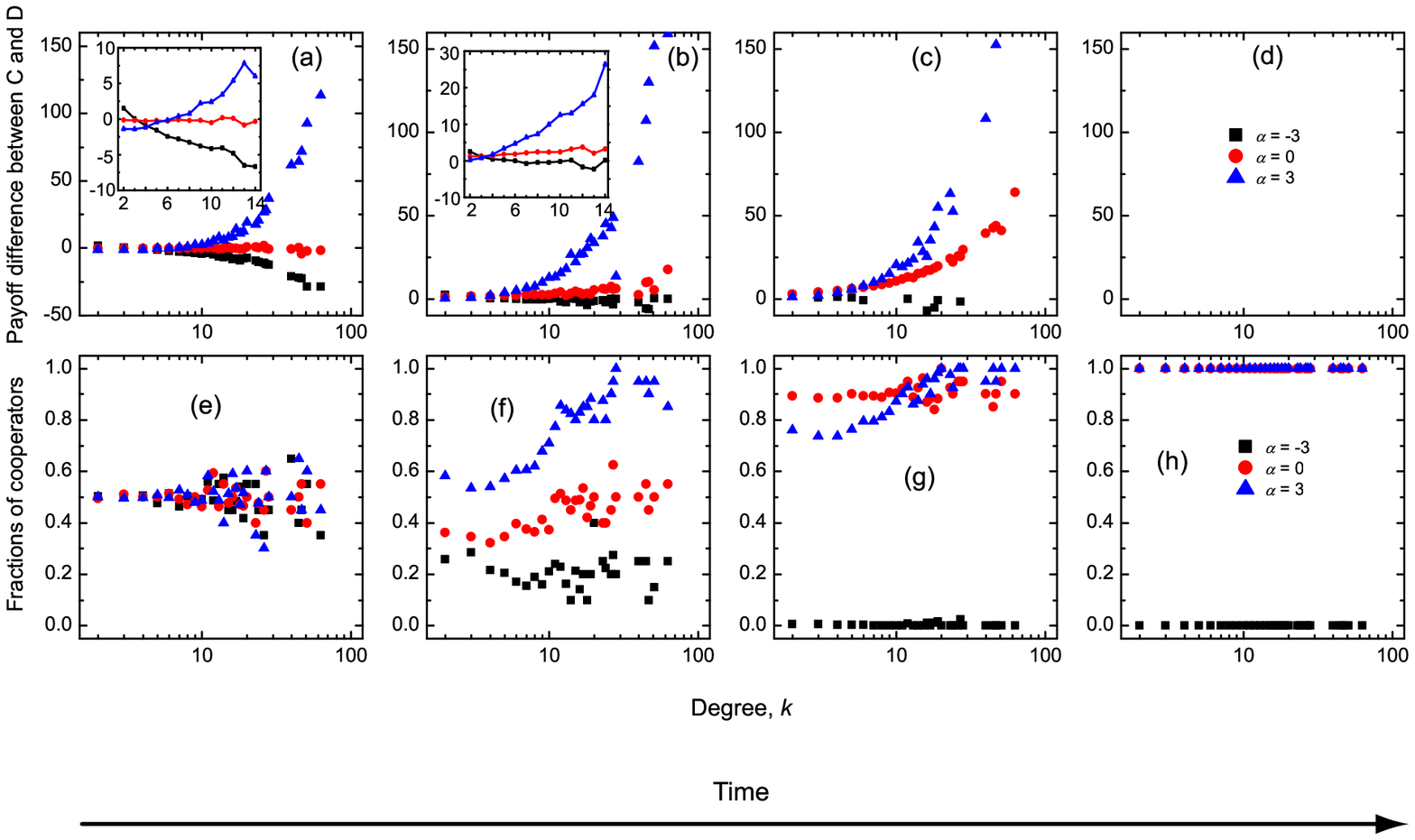}
\end{center}
\caption{The same results as presented in Fig.~\ref{fig5}, as obtained for a larger enhancement factor but a smaller average amount of available incentives. Parameter values are $r=2$ and $\delta=0.3$.}
\label{fig6}
\end{figure}

It remains of interest to explain why the dependence on $\alpha$ disappears at intermediate
values of $\delta$ for $r=2$. For this purpose, we show in Fig.~\ref{fig6} the same quantities as in Fig.~\ref{fig5}, from where it follows that the results do not change for $\alpha \leq 0$. However, for $\alpha>0$ the differences are clearly inferable. For $r=2$ the high-degree cooperators can obtain a positive payoff, and naturally they then have a higher payoff than high-degree defectors, because the latter receive ample negative incentives from the
institutional punishment pool [Fig.~\ref{fig6} (a)-(c)]. Cooperators can therefore occupy high-degree nodes and from there spread across the whole population [Fig.~\ref{fig6} (g)], and this the more effectively the higher the value of $\alpha$.

\subsection*{Institutional rewarding with degree-normalized payoffs}
\begin{figure}
\begin{center}
\includegraphics[width=12cm]{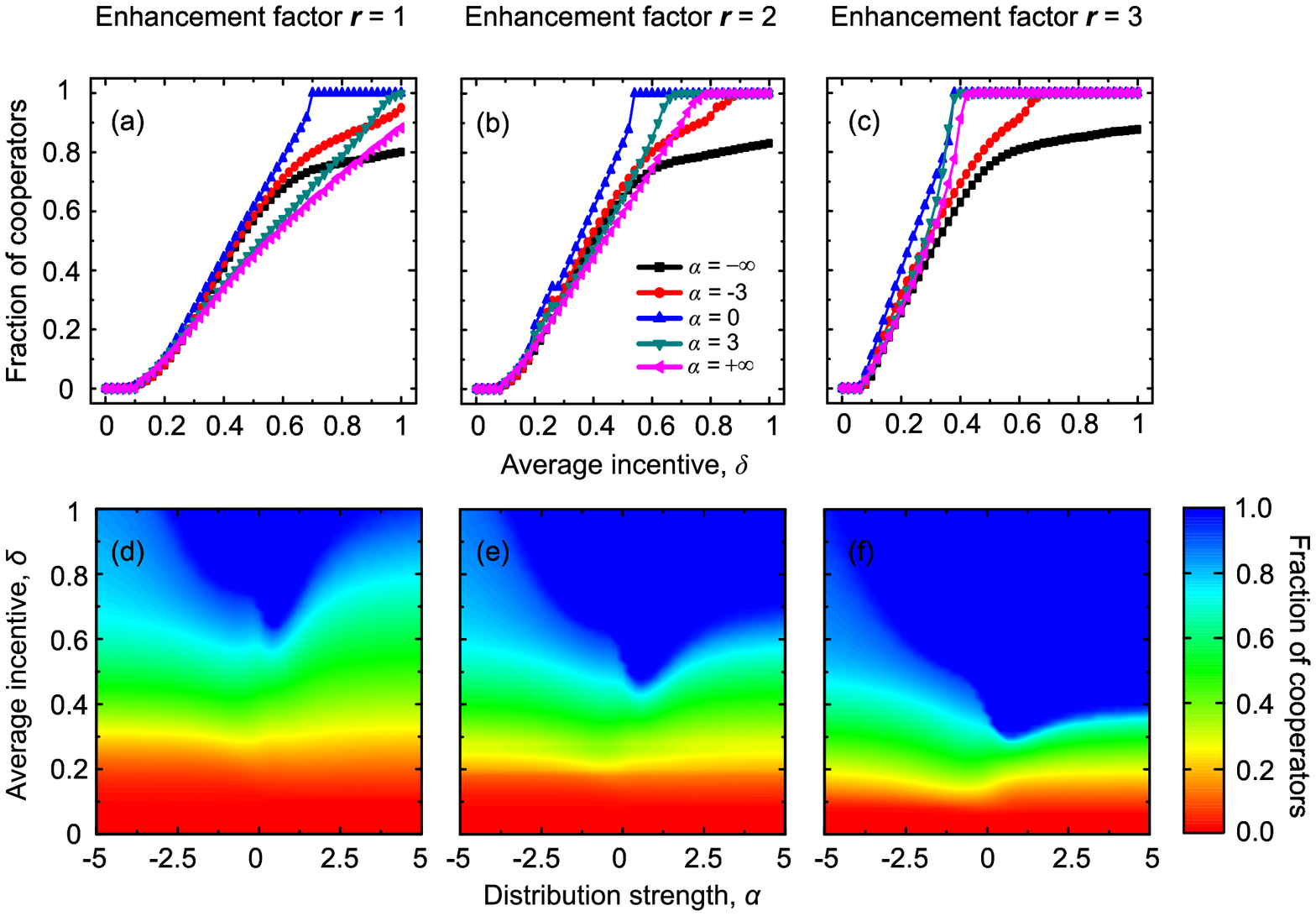}
\end{center}
\caption{Evolution of cooperation with institutional rewarding and degree-normalized payoffs. Top row depicts the stationary fraction of cooperators as a function of the average amount of available incentives $\delta$ for different values of the distribution strength $\alpha$. The enhancement factor is $r=1$ in (a), $r=2$ in (b), and $r=3$ in (c). Bottom row depicts the contour plot of the fraction of cooperators as a function of $\alpha$ and $\delta$, as obtained for the enhancement factor $r=1$ in (d), $r=2$ in (e), and $r=3$ in (f).}
\label{fig7}
\end{figure}

From here onwards we turn to considering degree-normalized payoffs, which can have important negative consequences for the evolution of public cooperation in heterogeneous environments if compared to absolute payoffs \cite{masuda_prsb07, tomassini_ijmpc07, wu_zx_pa07, szolnoki_pa08, maciejewski_ploscb14}. As shown in Fig.~\ref{fig7}, the fraction of cooperators unsurprisingly increases with increasing $\delta$ for various distribution strength $\alpha$ (top panels). Furthermore, at small values of $r$, for small values of $\delta$ defectors always dominate, regardless of the value of $\alpha$, while for intermediate values of $\delta$ cooperators recover gradually as $\alpha$ increases. For high $\delta$ values there exists an intermediate close-to-zero value of $\alpha$ that maximizes the stationary fraction of cooperators [Fig.~\ref{fig7}(a) and (d)]. When the enhancement factor is larger, the extent of the parameter region where the nonmonotonous phenomenon can be observed decreased. Instead, for high values of $\delta$, the cooperation level increases with increasing values of $\alpha$ and the area of complete cooperator dominance increases as well [Fig.~\ref{fig7}(e) and (f)]. If comparing the results presented in Fig.~\ref{fig7} with those presented in Fig.~\ref{fig1}, we find that the nonmonotonous phenomenon still exists, and it can appear even at larger values of $r$ because of the consideration of degree-normalized payoffs. In general, however, the explanation of these results and the evolutionary mechanisms behind are the same as those described when considering institutional rewarding with absolute payoffs.

\subsection*{Institutional punishment with degree-normalized payoffs}
\begin{figure}
\begin{center}
\includegraphics[width=12cm]{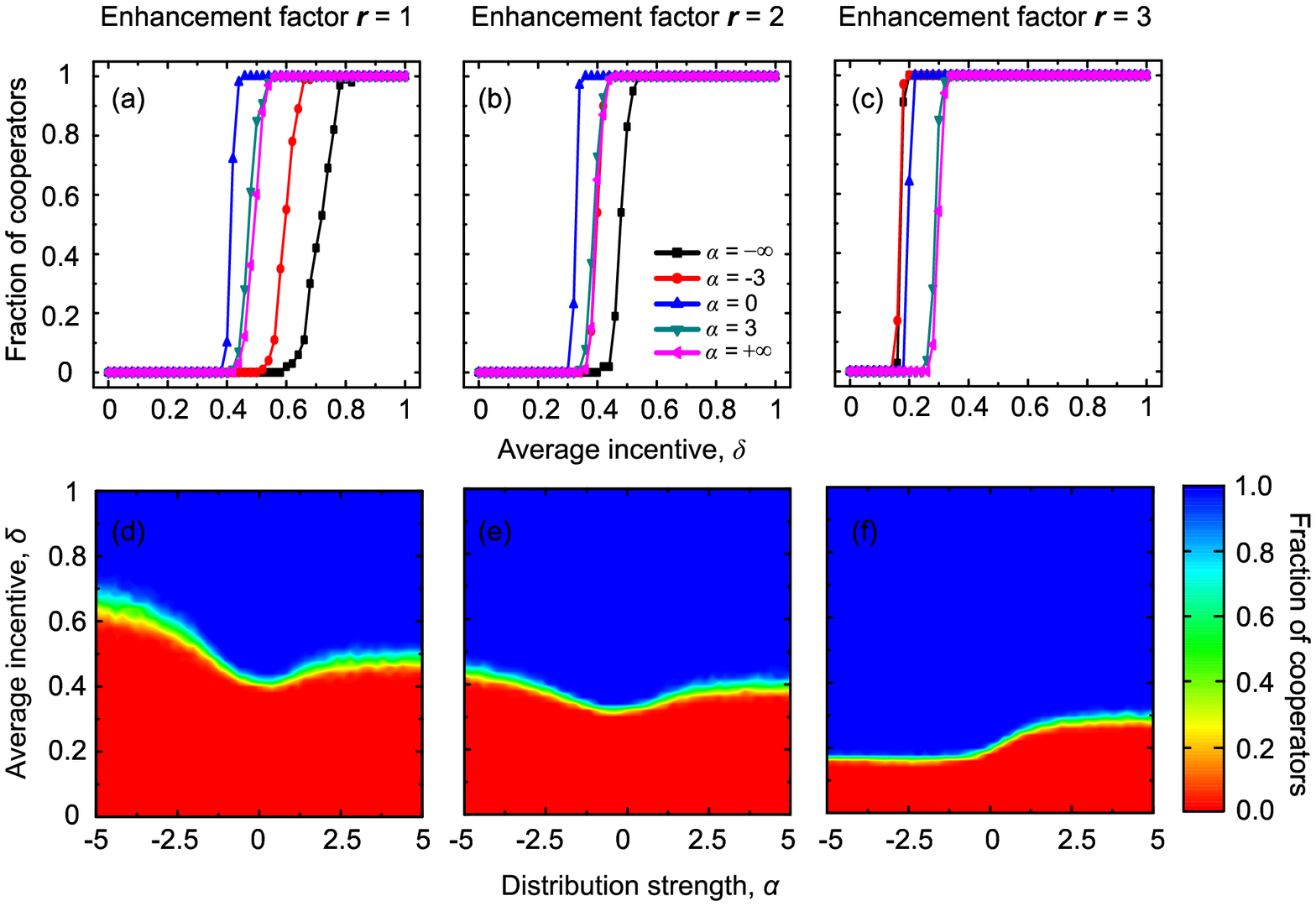}
\end{center}
\caption{Evolution of cooperation with institutional punishment and degree-normalized payoffs. Top row depicts the stationary fraction of cooperators as a function of the average amount of available incentives $\delta$ for different values of the distribution strength $\alpha$. The enhancement factor is $r=1$ in (a), $r=2$ in (b), and $r=3$ in (c). Bottom row depicts the contour plot of the fraction of cooperators as a function of $\alpha$ and $\delta$, as obtained for the enhancement factor $r=1$ in (d), $r=2$ in (e), and $r=3$ in (f).}
\label{fig8}
\end{figure}

Lastly, we consider institutional punishment with degree-normalized payoffs. From the results presented in Fig.~\ref{fig8} it follows that the stationary fraction of cooperators increases with increasing values of $\delta$, and this regardless of the value of $\alpha$ (top panels).
When the enhancement factor is small, we can see that the nonmonotonous dependence of the fraction
of cooperators on $\alpha$ exists at intermediate values of the average incentive $\delta$ [Fig.~\ref{fig8}(d)]. When the enhancement factor increases, this phenomenon still exists, but
the extent of the parameter region where the nonmonotonous dependence can be observed decreases [Fig.~\ref{fig8}(e)]. Surprisingly, when the enhancement factor increases further, the nonmonotonous dependence disappears. Instead, in a narrow region of intermediate $\delta$ values, the fraction of cooperators decreases with increasing values of $\alpha$. At the same time, the extent of the full cooperation area increases while the full defection region decreases in the considered $(\alpha, \delta)$ parameter space [Fig.~\ref{fig8}(f)].

While the underlying mechanism for the nonmonotonous dependence on $\alpha$ is qualitatively identical to that reported before when considering institutional punishment with absolute payoffs, the decrease of the level of cooperation at intermediate values of $\delta$ and $r=3$ as $\alpha$ increases requires special attention. We note that when the value of $r$ is large, low-degree cooperators can still have a positive payoff. For negative
$\alpha$, these low-degree cooperators can even have the highest payoffs because of the consideration of degree-normalized payoffs and sufficiently high values of $\delta$ to weigh heavily on the defectors. Therefore, cooperators dominate on all low-degree
nodes and from there spread further across the whole network and rise to dominance. This atypical spreading is a unique consequence of the consideration of the optimal distribution of negative incentives from the punishment pool, and it highlights the importance of the parameter $\alpha$. For positive values of $\alpha$ namely, because most of the negative incentives are then assigned to high-degree defectors and there are only a few of those in the entire population \cite{barabasi_s99}, the majority of low-degree defectors is not punished at all. The previously described spreading of cooperators from the low-degree nodes outwards is therefore impaired, which ultimately results in an overall lower stationary fraction of cooperators. Instead of cooperation, for larger values of $\alpha$ the low-degree nodes ``emit'' defection throughout the population.

\section*{Discussion}
To summarize, we have studied how to best distribute limited institutional incentives in order to maximize public cooperation on scale-free networks. We have considered both institutional rewarding of cooperators and institutional punishment of defectors, and we have also distinguished between absolute and degree-normalized payoffs. Our key assumptions was that, since in heterogeneous environments players have a different number of partners, the incentives ought to be distributed by taking this into account. This would be in agreement with the established importance of degree heterogeneity for cooperation in evolutionary games \cite{santos_pnas06, santos_n08, santos_jtb12}. Traditionally, however, previous research has considered the limited budged be distributed equally among all the potential recipients of the incentives, irrespective of the players status and influence within the network. Accordingly, how to distribute the incentives to optimize public cooperation was an important open problem.

We have found interesting solutions on how to optimally distribute
the incentives based on each player's social influence level, the
proxy for which are the number of social ties the players have
within the interaction network. We have shown that sharing the
incentives equally among all regardless of status is optimal only if
the social dilemma is strong and the propensity to contribute to the
common pool is thus weak, and if in addition the available amount of
incentives is intermediate. This result is valid for both
institutional punishment and institutional rewarding, and it does
not depend on whether absolute or degree-normalized payoffs count
towards evolutionary fitness. However, if the environment already
favors cooperative behavior --- when the public goods game is
characterized with a high enhancement factor --- then it is best to
reward influential players more than low-degree players, and this
regardless of whether absolute or degree-normalized payoffs apply.
For institutional punishment, on the other hand, the solution of the
optimization problem depends on whether absolute or
degree-normalized payoffs are used. We have shown that
degree-normalized payoffs require high-degree nodes be punished more
lenient than low-degree nodes, while if absolute payoffs count, then
high-degree nodes should be punished stronger than low-degree nodes.
In general, rewarding influential cooperators strongly and punishing
auxiliary defectors leniently appears to be optimal for the
successful evolution of public cooperation.

In terms of solving actual common goods problems, our work might have merit in situations with strong diversity in roles and group sizes. One representative example of such a situation is climate change governance, where existing research has shown that local institutions are an effective way to promote the emergence of widespread cooperation \cite{vasconcelos_ncc13}. Since our results are derived not only from local institutions, but take into account also the heterogeneous interaction environment, they could offer further advice on how to arrive at globally acceptable climate policies \cite{vasconcelos_pnas14}.

While the evolution of institutions remains a puzzle
\cite{sigmund_n10}, their importance for enforcing socially
acceptable behavior in human societies can hardly be overstated.
Although institutionalized punishment appears to be prevailing,
recent research concerning the effectiveness of punishment, for
example related to antisocial punishment \cite{herrmann_s08,
rand_nc11}, reciprocity \cite{ohtsuki_n09}, and reward
\cite{rand_s09, hilbe_prsb10}, is questioning the aptness of
sanctioning for elevating collaborative efforts and raising social
welfare. Indeed, although the majority of previous studies
addressing the ``stick versus carrot'' dilemma concluded that
punishment is more effective than reward in sustaining public
cooperation \cite{sigmund_tee07}, evidence suggesting that rewards
may be as effective as punishment and lead to higher total earnings
without potential damage to reputation or fear from retaliation is
mounting \cite{dreber_n08, szolnoki_epl10, szolnoki_njp12}. In
particular, Rand and Nowak \cite{rand_nc11} argue convincingly that
healthy levels of cooperation are likelier to be achieved through
less destructive means. We hope that our study will prove to be
inspirational for further research aimed at discerning the
importance of positive and negative reciprocity for human
cooperation, as well as for looking closely at their correlated
effects \cite{szolnoki_prx13, chen_xj_submitted14}.

\section*{Methods}
We consider the evolutionary public goods game on the Barab{\'a}si-Albert scale-free network \cite{barabasi_s99, albert_rmp02}. Each player $x$ occupies one node of the network, and it can choose between cooperation $(s_x=1$) and defection ($s_x=0$) as the two competing strategies. To each public goods game cooperators contribute the cost $c=1$, while defectors contribute nothing. The payoff of player $x$ who is member in the group $G_y$, which is centered on player $y$, depends on the size of the group $k_y+1$ (here $k_y$ is also the degree of node $y$), on the number of cooperators $n_c$ in the group, and on the enhancement factor $r$. In addition to the payoffs stemming from the public goods game, each group receives institutional incentives $I_x=(k_x+1)\delta$ to be used either for rewarding cooperators or for punishing defectors, where $\delta$ is the average amount of available incentives.

When the incentives are used for rewarding, a cooperator $y$ with degree $k_y$ that is member in the group $G_x$ thus receives the payoff
$$P_{y,x}=\frac{rcn_c}{k_x+1}-c+\frac{s_yk_y^{\alpha}}{\sum_{z\in G_x} s_z
k_z^{\alpha}}I_x,$$ while a defector in the same group receives $$P_{y,x}=\frac{rcn_c}{k_x+1},$$
where $\alpha$ is the distribution strength. According to the definition of the payoffs, for $\alpha>0$ high-degree nodes obtain larger rewards than low-degree nodes, while for $\alpha<0$ low-degree nodes receive a larger share from the incentive pool.

If the incentives are used for punishing defectors rather than rewarding cooperators, then a cooperator $y$ with degree $k_y$ that is member in the group $G_x$ receives $$P_{y,x}=\frac{rcn_c}{k_x+1}-c,$$ while a defector in the same group receives $$P_{y,x}=\frac{rcn_c}{k_x+1}-\frac{(1-s_y)k_y^{\alpha}}{\sum_{z\in G_x}
(1-s_z) k_z^{\alpha}}I_x.$$ As by institutionalized rewarding, here to $\alpha>0$ implies high-degree nodes are punished stronger than low-degree nodes, and vice versa for $\alpha<0$.

Each player $x$ participates in $k_x+1$ public goods games, which are staged in groups that are centered on player $x$ itself and on its $k_x$ neighbors, respectively. The total payoff player $x$ obtains is thus $P_x=\sum_{y\in G_x}P_{x,y}$. After playing the games, a player is allowed to learn from one of its randomly chosen neighbors $y$ and update its strategy accordingly. The probability of strategy change is given by the Fermi function \cite{szabo_pre98, szabo_pr07}
\begin{equation}
\label{eq.1}f = \frac{1}{{1 + \exp [({P_x} - {P_y})/K]}},
\end{equation}
if we assume that absolute payoff are considered. However, previous research has emphasized also the importance of degree-normalized payoffs \cite{masuda_prsb07, tomassini_ijmpc07, wu_zx_pa07, szolnoki_pa08, maciejewski_ploscb14}, in which case the probability of strategy change is
\begin{equation}
\label{eq.2}f = \frac{1}{{1 + \exp [({P_x/k_x} - {P_y/k_y})/K]}}.
\end{equation}
We consider both absolute (Eq.~\ref{eq.1}) and degree-normalized (Eq.~\ref{eq.2}) payoffs to be representative for the evolutionary fitness of individual players. Especially for institutional punishment, the solution of the considered optimization problem depends significantly on this difference. Without losing generality we set the uncertainly in the strategy adoption process to $K=0.1$ \cite{szolnoki_pre09c}, so that it is very likely that the better performing players will be imitated, although it is also possible that players will occasionally learn from those performing worse.

\section*{Acknowledgments}
This research was supported by the National Natural Science
Foundation of China (Grant 11161011) and the Slovenian Research
Agency (Grants J1-4055 and P5-0027). X. C. would like to thank Ulf
Dieckmann for helpful discussion.


\clearpage

\end{document}